\documentclass[12pt]{article}
\usepackage{epsfig, graphicx}
\usepackage{amsmath}
\usepackage{latexsym}
\usepackage{amsfonts}
\usepackage{amstext}
\newcommand{\beq}{\begin{eqnarray}}
\newcommand{\eeq}{\end{eqnarray}}
\newcommand{\beqq}{\begin{eqnarray*}}
\newcommand{\eeqq}{\end{eqnarray*}}
\newcommand{\p}{\partial}

\newcommand{\x}{\mbox{\boldmath$x$}}

\newcommand{\y}{\mbox{\boldmath$y$}}

\begin{document}
\pagestyle{plain}
\begin{center} {\large \textbf{{The probability of an encounter of two Brownian particles before escape}}\\[5mm]
D. Holcman \footnote{D\'epartment of Computational Biology, Ecole
Normale Sup\'erieure, 46 rue d'Ulm 75005 Paris, France. This
research is supported by the program ERC-Starting Grant in
Mathematics.} and  I. Kupka
\footnote{D\'epartement de math\'ematique, Paris VI, 174 rue du
Chevaleret, 75013 Paris, France and Department of Computational
Biology, Ecole Normale Sup\'erieure, 46 rue d'Ulm 75005 Paris,
France. }}
\end{center}
\date{}
\begin{abstract}
We study the probability of two Brownian particles to meet before
one of them exits a finite interval. We obtain an explicit
expression for the probability as a function of the initial distance
of the two particles using the Weierstrass elliptic function. We
also find the law of the meeting location. Brownian simulations show
the accuracy of our analysis. Finally, we discuss some applications
to the probability that a double strand DNA break repairs in
confined environments.
\end{abstract}

keywords: Brownian motion, conformal mapping, Weierstrass function.

\subsection*{Introduction}

The problem of coalescence and clustering in an open space has been
considered by Chandrasekar \cite{Chandrasekar} (see also
\cite{Redner} for a review). Little work has been dedicated to the case of a
competition between the coalescence of Brownian independent
particles and the possible escape at the boundary of a domain where
the particles are absorbed. This situation is however reminiscent of
many biophysical problems. For example, the probability of a correct
repair of a broken DNA molecule inside the nucleus. One mode of
repair is known as non-homologous end-joining \cite{NHEJ}, which
depends crucially on the initial distance between the two free DNA
strands: either the branches meet or one of them can curl up before
and then the probability to connect is very low (almost zero). Here
we consider the drastic simplification that the motion of the DNA
molecule tip can be approximated as a one dimensional Brownian
motion (see the discussion).

We consider the motion of two independent Brownian particles
$X_1(t),X_2(t)$ inside an interval $[a,b], (a<b)$ with the following
rules: when the two particles meet, they coalesce into a single one
subjected to a Brownian motion. We compute the probability $P_M$
that the two particles meet before one of them hits the boundary of
the interval and obtain an explicit expression for the probability
of the two particles clustering, as a function of the initial
position. When the initial points are $a<x_1<x_2<b$, we obtain that
\beq
P_M(x_1,x_2)=\frac{-2}{\pi }\Im m \log \mathfrak{P}\left(
\frac{\omega (Z-a)}{L\sqrt{8} }\right),
\eeq
where $\Im m$ is the imaginary part, $\mathfrak{P}$ is the
Weierstrass elliptic function defined by equation (\ref{weier}),
$L=b-a$, $Z=x_2+\sqrt{-1}x_1$, and
\beq \label{omegaa}
\omega=\int_{1}^{+\infty
}\frac{dx}{\left[ x(x-1)\right] ^{\frac{3}{4}}}=5.244115106
\eeq
is a universal number defined  by an elliptic integral. We further
obtain the probability distribution of their meeting point. Finally,
the analytical formulas are compared with Brownian simulations,
where we gain the information about the variance. The role of the
Weierstrass elliptic function is quite surprising here and really
comes from the method of conformal mapping. We wonder if our result
can be recovered from elementary probability arguments. A Brownian
interpretation of an elliptic integral was given in \cite{Yor}.


\subsection*{Formulation}
The dynamics of each particle is given for $i=1,2$ by
\beq
d{X}_i =\sqrt{2D_f}d{w_i}
\eeq
where $D_f$ is the diffusion constant and $w_1,w_2$ are two Brownian
motions of unit variance. We are interested in the probability $P_M$
that the two particles meet before one of them exits the interval
$[a,b]$. If we consider the two random times
\beqq
\tau_1 &=&\inf \{ t>0, X_1(t)=a \hbox{ or } X_2(t)=b, X_1(0)=x_1 \hbox{ and } X_2(0)=x_2, x_1<x_2
\}\\
\tau_2 &=&\inf \{ t>0, X_1(t)=X_2(t), X_1(0)=x_1 \hbox{ and } X_2(0)=x_2, x_1<x_2
\},
\eeqq
then for $\x=(x_1,x_2)$, the probability
\beq
P_M(\x) =Pr\{\tau_2<\tau_1| \x \}
\eeq
satisfies the Laplace equation
\beq \label{1}
\Delta P_M(\x) &=&0 \hbox{ for } \x \in T\\
P_M(\x) &=&1 \hbox{ for } \x \in  D \nonumber\\
P_M(\x) &=&0 \hbox{ for } \x  \in \p T-D \nonumber
\eeq
where T is a right-angled triangle with vertices $a,b,b+a\sqrt{-1}.$
$D$ is the side joining $a$ to $b+a \sqrt{-1}.$ Indeed,
\beq
P_M(\x)=Pr(X(\tau)=\y \in T|X(0)=\x) =\int_{T} G(\x,\y) dS_{\y}
\eeq
where $\tau$ is the first exit time and the Green function $G$ is
solution of (see \cite{book})
\beq \label{2}
\Delta G(\x,\y) &=&0  \hbox{ for } \x \in T\\
G(\x,\y)  &=&\delta_{\y}(\x)  \hbox{ for } \x \in  D \nonumber\\
G(\x,\y)  &=&0 \hbox{ for } \x  \in \p T-D. \nonumber
\eeq
$G(\x,\y)$ is the probability density function to exit in $\y \in D$
when the particle starts initially in $x$ (see also ch. 15, p.192 of
reference
\cite{Karlin} for another proof). We shall derive an explicit expression of the encounter probability
$P$. To solve equation, we shall use the invert of a
Schwarz-Christoffel mapping to map the triangle into the upper
complex half-plan $H$. By using the explicit the solution of the
Laplace equation in $H$, we will find the solution of (\ref{1}). It
turns out that the Schwarz-Christoffel mapping  of interest is a
Weierstrass function.

\subsection*{Analytical derivation of the encounter probability }
It will be convenient to do all our computations for the choice
$a=0,b=\omega,$ where $\omega$ is defined in \ref{omegaa}.
Note that $\omega $ and $\omega \sqrt{-1}$ are a pair of fundamental periods
for the Weierstrass $\mathfrak{P}$ function with parameters $g_{2}=1$ and $%
g_{3}=0,$ \cite{Lawden} defined by
\beq \label{weier}
\mathfrak{P}'^2=4\mathfrak{P}^3-g_2 \mathfrak{P}-g_3,
\eeq
a function we are going to use in the following. It is a matter of
a dilation to deduce the results for the case of general $ a,b $
from our special case.

\begin{figure}
{\includegraphics[width=.53 \textwidth]{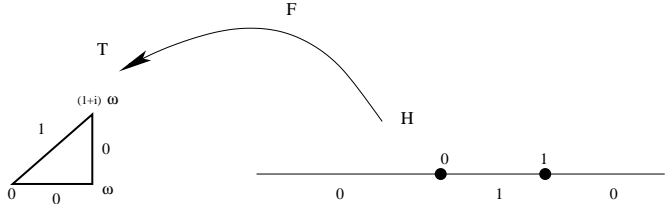}}
\caption{\small Transformation F from the triangle to the upper complex plane. We position the boundary
condition for the Laplace equation on the associated part of the
boundary} \label{encounter}
 \end{figure}

In the spaces $H=\{z\in \mathbb{C}| \Im mz >0\}$ and $ \overline{H}=
\{z \in \mathbb{C}| \Im m z \geq 0\},$  we consider  $f:H\rightarrow \mathbb{C}$
to be the branch of $\left[ z(z-1)\right]^{\frac{3}{4}}$ defined as follows: for $z\in H$, we set
\beq
z=r_{0}\exp \left( \theta _{0}\sqrt{-1}\right) \hbox{ and } z-1=r_{1}\exp \left( \theta _{1}\sqrt{-1}\right)
\eeq
where $r_{0}=|z|,$ $r_{1}=|z-1|,$ $ 0\leq \theta _{0},\theta
_{1}\leq \pi .$ In that case,
\beq
f(z)=\left( r_{0}r_{1}\right) ^{\frac{3%
}{4}}\exp \left[ \left( \frac{\theta _{0}+\theta _{1}}{4}\right) 3\sqrt{-1}%
\right].
\eeq
f has a continuous extension to $\overline{H}:$%
\beq
f(x)=
\left\{ \begin{array}{cc}
\left[ x(x-1)\right]^{\frac{3}{4}} &  \text{ if }x\geq 1\\
-\sqrt{-1}\left[ \left\vert x(x-1)\right\vert\right]^{\frac{3}{4}} &\text{ if } x\leq 0\\
-\left( \frac{1+\sqrt{-1}}{2}\right) \left[ x(1-x)\right]
^{\frac{3}{4}} &\text{ if } 0\leq x \leq 1
\end{array}\right.
\eeq
We shall now define $F:\overline{H}\rightarrow \mathbb{C}$ by:
\beq \label{eqq}
F(\zeta )=\int\limits_{1}^{\zeta }\frac{dz}{f(z)}
\eeq
The Schwarz'reflection lemma shows that $F$ is a conformal mapping
of $\overline{H}$ onto the triangle $T$ in $\mathbb{C}$ having as vertices $0,\omega ,\left( 1+\sqrt{-1}%
\right) \omega $ where $\omega =\int\limits_{1}^{+\infty }f(x)dx$ (see figure
\ref{encounter}). $F$ maps $1,\infty ,0$ onto $0,\omega ,\left( 1+\sqrt{-1}\right)
\omega $ respectively and the half-line $\left[ 1,+\infty \right] $
onto the segment $
\left[ 0,\omega \right] $ of the real axis, the half line $\left[ -\infty ,0%
\right] $ onto $\{\omega +t\omega \sqrt{-1}|0\leq t\leq 1\}$ and the segment
$\left[ 0,1\right] $ of the real axis onto $\{\left(
1+\sqrt{-1}\right) (1-t)\omega |0\leq t\leq 1\}.$ Moreover $F$ is
conformal on $H.$

To compute the function $F$ given by (\ref{eqq}), or more precisely
its inverse we introduce the following transformation: $z\in
H\rightarrow p=\varphi (z)\in \mathbb{C}$,
\beq
\varphi (z)=\sqrt{\frac{z}{4(z-1)}},
\eeq
where the square root is the one such that
${\Re}p\geq 0.$ With the above notations, $p=\frac{1}{2}\sqrt{\frac{r_{0}}{r_{1}%
}}\exp \left[ \frac{\theta _{0}-\theta _{1}}{2}\sqrt{-1}\right] $.
$\varphi $ is a homeomorphism of $\overline{H}$ onto the quadrant
$\overline{Q}=\{p\in
\mathbb{C}| \Re e p\geq 0,\Im m p\leq 0\}$ and $\varphi $ is conformal
on $Q=\{p\in \mathbb{C}|\Re e p>0,\Im p<0\}.$ On the space $H$,
\beq
f(z)dz=\varphi ^{\ast }(g(p)dp)  \label{un}
\eeq
where
\[
g(p)=\frac{\sqrt{8}}{\sqrt{4p^{3}-p}}
\]
and $\sqrt{4p^{3}-p}$ is the branch in $Q$ which is real positive on $]%
\frac{1}{2},+\infty \lbrack .$ $\varphi $ maps $[1,+\infty ]$ onto $[+\infty ,\frac{1}{2}],$ $[0,1]$ onto $-%
\sqrt{-1}[0,+\infty ]$ and $[-\infty ,0]$ onto $[\frac{1}{2},0].$

Relation (\ref{un}) in (\ref{eqq}) implies that:%
\beq
F(\zeta )=\int_{\varphi (\zeta )}^{+\infty }g(p)dp
=\sqrt{8}\int_{\varphi (\zeta )}^{+\infty
}\frac{1}{\sqrt{4p^{3}-p}}dp.
\label{deux}
\eeq
We recall that the Weierstrass elliptic function
\cite{Lawden},$\mathfrak{P}$ with $g_{2}=1,g_{3}=0$ is defined as the
$\zeta=\mathfrak{P}(\xi)$ such that
\beq
\xi=\int_{\zeta}^{+\infty
}\frac{1}{\sqrt{4p^{3}-p}}dp.
\eeq
Relation (\ref{deux}) implies that:%
\beq
\varphi (\zeta )=\mathfrak{P(}\frac{F(\zeta )}{\sqrt{8}}\mathfrak{)}
\label{trois}
\eeq
Equivalently, (\ref{trois}) implies that:%
\[
\zeta =\frac{\left( 4\mathfrak{P(}\frac{F(\zeta )}{\sqrt{8}}\mathfrak{)}%
\right) ^{2}}{\left( 4\mathfrak{P(}\frac{F(\zeta )}{\sqrt{8}}\mathfrak{)}%
\right) ^{2}-1}
\]
Hence we have for $Z\in T:$%
\[
F^{-1}(Z)=\frac{\left(
4\mathfrak{P(}\frac{Z}{\sqrt{8}}\mathfrak{)}\right) ^{2}}{\left(
4\mathfrak{P(}\frac{Z}{\sqrt{8}}\mathfrak{)}\right) ^{2}-1}.
\]
To compute (\ref{1}), we shall find a harmonic function $u$ with the
boundary condition
\begin{eqnarray*}
u|]0,\omega ]\cup \{\omega +(t\omega \sqrt{-1}|0 &\leq &t<1\}=0 \\
u|\{\left( 1+\sqrt{-1}\right) t\omega |-1 &<&t<0\}=1.
\end{eqnarray*}
This harmonic function can be expressed using the Harmonic function
$v$ in $\overline{H}$ with the boundary conditions
\beq
v= \left\{ \begin{array}{ccc} 0 & \text{ on }[-\infty ,0[\cup
]1,+\infty ]\\ & \\1 & \text{ on }\left[ 0,1\right]
\end{array}\right.
\eeq
It is given for $Z\in T$ by:%
\[
u(Z)=v(F^{-1}(Z))=v\left( \frac{\left( 4\mathfrak{P(}\frac{Z}{\sqrt{8}}%
\mathfrak{)}\right) ^{2}}{\left( 4\mathfrak{P(}\frac{Z}{\sqrt{8}}\mathfrak{)}%
\right) ^{2}-1}\right)
\]
Actually, the function $v$ is given for $z\in H:$ by
\beq
v(z)=\frac{1}{\pi }\Im m \log \left(\frac{z-1}{z} \right)
\eeq
where $\log \frac{z-1}{z}$ is the branch on $\overline{H}$
$-\{0,1\}$ that is real for $z\in ]1,+\infty \lbrack $ i.e.:
\beq
\log \frac{z-1}{z}=\log \frac{r_{0}}{r_{1}}+\sqrt{-1}(\theta _{1}-\theta
_{0})
\eeq
Finally, we obtain for $Z\in T$ the expression
\beq
u(Z)=\frac{-2}{\pi }\Im m \log
\mathfrak{P(}\frac{Z}{\sqrt{8}}\mathfrak{)}
\eeq
where $\log $ is the branch on $\overline{H}$ $-\{0\}$ which is real on $%
]0,+\infty \lbrack .$ This result can be seen as a generalization of
Spitzer's law concerning the winding of a two dimensional Brownian
motion \cite{Spitzer}.

\subsection*{Comparison with Simulations}
\begin{figure}
{\includegraphics[width=.53 \textwidth]{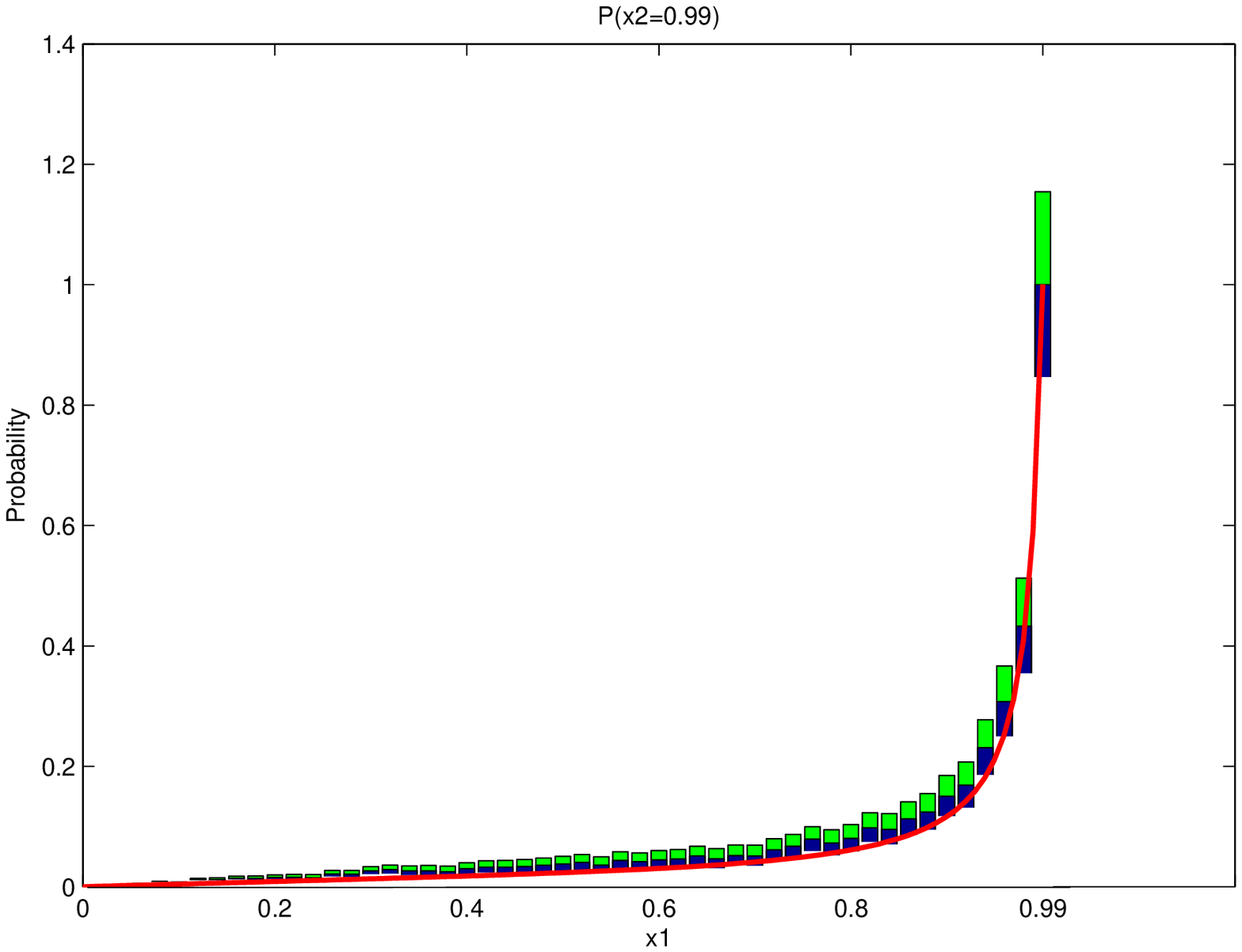}}
\includegraphics[width=.53 \textwidth]{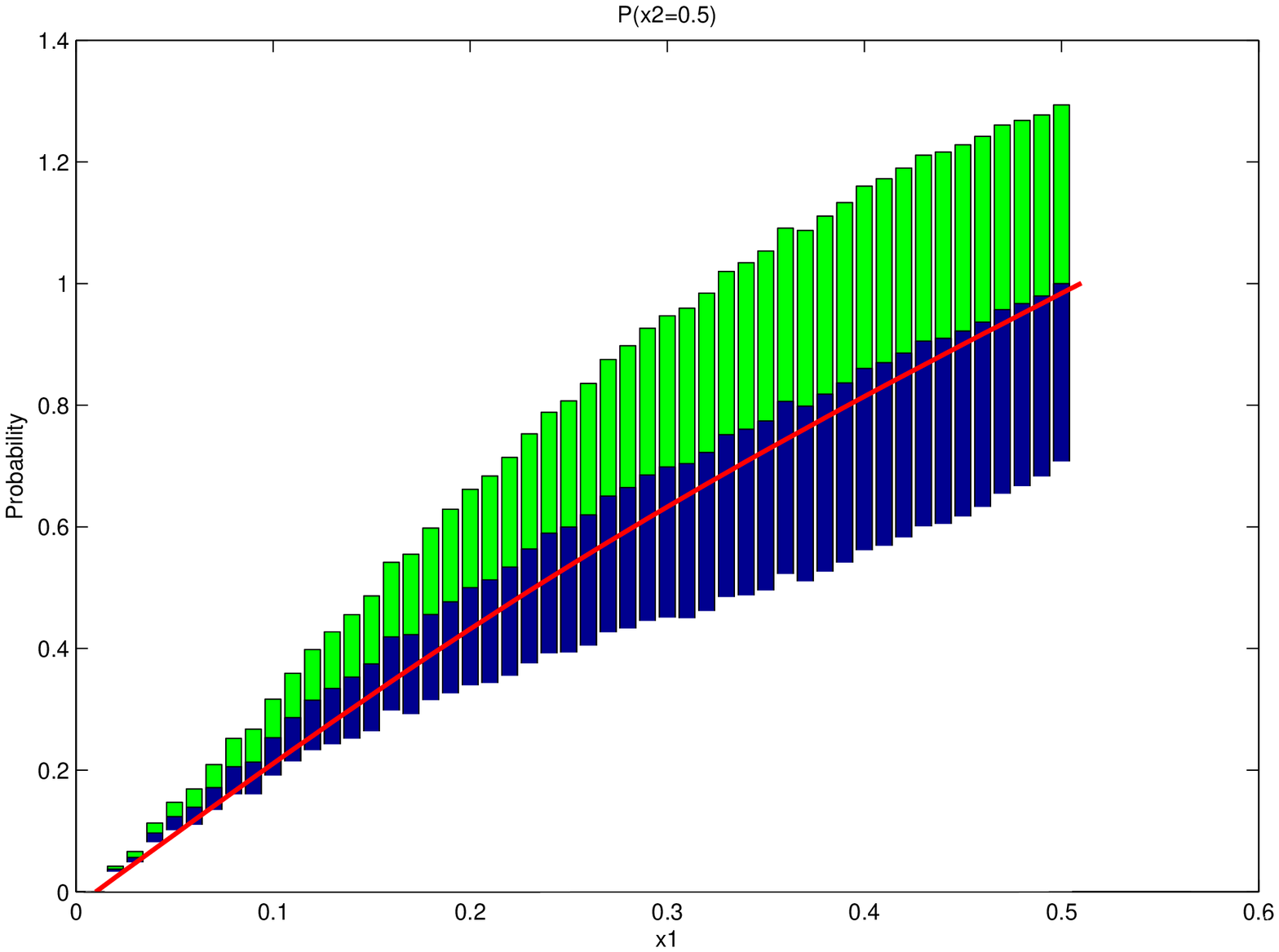}
\caption{\small Probability $P_M$ for two Brownian particles to meet
 before one of them escapes the interval $[0,1]$. We  compare the analytical solution Eq. \ref{1} with
 Brownian simulations. For each position, we averaged 2000 realizations. The variance is presented as error bars.
 On the right, we start at a middle point of the interval $(x_2=0.5)$ and plot the probability as a
 function of $x_1$ (with $x_1 <x_2$ for the second point, while on the left, we chose for $x_2$ a point very close to the boundary
 $(x_2=0.99)$. The effect of the boundary layer appears clearly for $(x_2=0.99)$.}
\label{schematic}
 \end{figure}

When the initial positions of the Brownian particles are $0\leq
x_1<x_2 \leq 1$, the scaled meeting probability is given by
\beq
P_M(x_1,x_2)=\frac{-2}{\pi }\Im m \log
\mathfrak{P(}\omega (\frac{x_2+\sqrt{-1}x_1}{\sqrt{8}})\mathfrak{)},
\eeq
where $\omega$ is defined in equation \ref{omegaa}.
In figure \ref{schematic},  we present the graph of the probability
$P_M$ of forming a cluster, plotted as a function of the initial
positions $x_1$ with the restriction that $x_1<x_2$. We present two
simulations: in the first one, we fix the point $x_2$ at the middle
of the interval $(x_2=0.5)$ and the other graph is obtained for a
point $x_2$ chosen very close to the boundary $(x_2=0.99)$. As can
be observed, the shape of the encounter probability changes
drastically. We have superimposed in figure \ref{schematic}, the
Brownian simulations (mean and variance) with the analytical
solution.

We remark that the probability to meet does not depend on the
diffusion constant.

\subsection*{The position of encounter}
To satisfy our last curiosity,  we finish with the computation of
the probability $p(Z;E),Z=x_{2}+\sqrt{-1}x_{1}$ for the two
particles, starting at positions $(x_{1},x_{2})$, $0\leq
x_{1},x_{2}\leq \omega "$ to coalesce in a measurable subset $E$ of
$[0,\omega].$ Given $E \in \lbrack 0,\omega],$ the function $Z\in T
\rightarrow$ $p(Z;E)\in \lbrack 0,1]$ is harmonic in the interior of
$T$ and
\beq
p(Z;E)=
\left\{ \begin{array}{cc}
0 &  \text{ if } z \in [0,\omega
]-E\\
1 & \text{ if }  z \in E
\end{array}\right.
\eeq
Using the conformal transformation $T\rightarrow \overline{H},$
\beq
Z\rightarrow z=\psi (Z)=\frac{4\left(
\mathfrak{P(}\frac{Z}{\sqrt{8}}
\mathfrak{)}\right) ^{2}}{4\left( \mathfrak{P(}\frac{Z}{\sqrt{8}}\mathfrak{)}%
\right) ^{2}-1},
\eeq
it is sufficient to compute for any measurable subset $M$ of $ [0,
1]$, the function: $z\in \overline{H}\rightarrow P(z;M)\in
\lbrack 0,1] $ having the following properties:

\begin{enumerate}
\item P is harmonic in $H$.
\item
\beq
P(z;M)=
\left\{ \begin{array}{cc}
0 &  \text{ if } z \in [0,1]-M\\
1 &  \text{ if } z \in M
\end{array}\right.
\eeq
\item for any $z\in H,$ $M\in \mathfrak{M(}[0,1]\mathfrak{)\rightarrow }P(z;M)$ is a measure on
the $\sigma $-algebra $\mathfrak{M(}[0,1]\mathfrak{)}$ of measurable
subsets of $[0,1].$ Then,
\[
p(Z;A)=P(\psi (Z),\psi (A)).
\]
\end{enumerate}
We shall remark that condition $3)$ above shows that to determine
$P$, it is sufficient to compute $P(z,.)$ for $M$ a closed interval
$[\alpha,\beta]$ of $[0,1].$ This is similar to the determination of
the function $v$ above. Hence,
\beq
P(z;[\alpha,\beta])=\frac{1}{\pi }\Im m \log
\frac{z-\beta}{z-\alpha}.
\eeq
This expression shows that for $z\in H,$ $M\in \mathfrak{M(}[0,1]\mathfrak{%
)\rightarrow }P(z;M)$ has a density $D(z;\alpha ),z\in H,\alpha \in
\lbrack
0,1],$ with respect to the Lebesgue measure on $[0,1]:$%
\[
D(z;\alpha )=\frac{\partial P}{\partial
\alpha}(z;[\alpha,\beta])=-\frac{1}{\pi }\Im m\frac{1}{z-\alpha }.
\]
Hence $p(Z;E)$ has a density with respect to the arc length on the segment $%
\{\left( 1+\sqrt{-1}\right) \left( 1-t\right) \omega |0\leq t\leq 1\},$
which, for $A\in \{\left( 1+\sqrt{-1}\right) \left( 1-t\right)
\omega |0\leq
t\leq 1\},$ is equal to :%
\beq \label{density}
d(Z;A)=D(\psi (Z),\psi (A))\left\vert \frac{d\psi
}{dZ}(A)\right\vert
\eeq
Recall that $\psi $ maps the segment $\{\left( 1+\sqrt{-1}\right)
\left(
1-t\right) \omega |0\leq t\leq 1\}$ diffeomorphically onto the interval $%
[0,1].$
\beq
\frac{1}{\psi (Z)-\psi (A)}=\frac{\left( \mathfrak{P}^{^{\prime }}\mathfrak{(%
}\frac{Z}{\sqrt{8}}\mathfrak{)P}^{^{\prime }}\mathfrak{(}\frac{A}{\sqrt{8}}%
\mathfrak{)}\right) ^{2}}{\left( \mathfrak{P(}\frac{Z}{\sqrt{8}}\mathfrak{)P}%
^{^{\prime }}\mathfrak{(}\frac{A}{\sqrt{8}}\mathfrak{)}\right)
^{2}-\left(
\mathfrak{P}^{^{\prime }}\mathfrak{(}\frac{Z}{\sqrt{8}}\mathfrak{)P(}\frac{A%
}{\sqrt{8}}\mathfrak{)}\right) ^{2}},
\eeq
where
\beq
\mathfrak{P}^{^{\prime }}\mathfrak{(}w\mathfrak{)=}\frac{d\mathfrak{P}}{dw}%
\mathfrak{(}w\mathfrak{)}
\eeq
and using the different equation satisfied by $\mathfrak{P}$, we
obtain that
\beq
\frac{d\psi }{dZ}=-\sqrt{8}\left( \frac{\mathfrak{P(}\frac{Z}{\sqrt{8}}%
\mathfrak{)}}{\mathfrak{P}^{^{\prime }}\mathfrak{(}\frac{Z}{\sqrt{8}}%
\mathfrak{)}}\right) ^{3}
\eeq
Finally the meeting density function (\ref{density}) is given
explicitly by
\beq
d(Z;A)=\frac{\sqrt{8}}{\pi }\left\vert \frac{\mathfrak{P}\left( \frac{A}{%
\sqrt{8}}\right) }{\mathfrak{P}^{\prime }\left( \frac{A}{\sqrt{8}}\right) }%
\right\vert ^{3} \Im m\frac{\left( \mathfrak{P}^{\prime }\left( \frac{Z}{%
\sqrt{8}}\right) \mathfrak{P}^{\prime }\left( \frac{A}{\sqrt{8}}\right)
\right) ^{2}}{\mathfrak{P}\left( \frac{Z}{\sqrt{8}}\right) \left( \mathfrak{P%
}^{\prime }\left( \frac{A}{\sqrt{8}}\right) \right) ^{2}-\left( \mathfrak{P}%
^{\prime }\left( \frac{Z}{\sqrt{8}}\right) \right) ^{2}\mathfrak{P}\left(
\frac{A}{\sqrt{8}}\right) }
\eeq
where $Z\in T$ and $A\in \{\omega (1-t)(1+\sqrt{-1})|0\leq t\leq
1\}$.
If $E$ is a measurable subset of the state space $[0,\omega ]$ then
the probability for two particles starting at position Z to meet in
$E$ is given by
\beq
p(Z;E)=\sqrt{2}\int_{E}d(Z;(a+a\sqrt{-1}))da,
\eeq
which conclude this part. To finish, we extend our formula for an
initial segment $[a,b]$. In that case, the probability to meet
before escape scales into:
\beq\label{eqp}
P(Z)=\frac{-2}{\pi }\Im m  \log \mathfrak{P}\left( \frac{\omega (Z-a)}{L\sqrt{8}%
}\right)
\eeq
where $Z\in T^{\prime }$ the triangle in $\mathbb{C}$ with vertices $%
a,b,b+(b-a)\sqrt{-1}$. In the limit of L large, using the double
pole expansion of $\mathfrak{P}$ at zero, for $\tilde Z$ in a
neighborhood of 0, we have
\beq
\mathfrak{P}(\tilde Z)=\frac{1}{\tilde Z^2}+O(\tilde Z),
\eeq
thus for large $L$, for $\tilde Z=\frac{\omega
(Z-a)}{L\sqrt{8}}\approx\frac{\omega Z}{L\sqrt{8}}$ equation
(\ref{eqp}) becomes
\beq
P(Z)\rightarrow_{L\rightarrow \infty}\frac{4}{\pi } \Im m \log
 \tilde Z=\frac{4}{\pi }\arctan \left(\frac{x_1}{x_2}\right).
\eeq
We shall remark that this law is twice the one of a Cauchy variable
at time 1, that is $Prob\{|C_1|<\frac{x_1}{x_2}\}$. This suggests
that this asymptotic result might be recovered by elementary
considerations on the Brownian motion. Finally, the meeting
probability density function at the point A is now given by:
\beqq
d(Z;A)=\frac{\omega \sqrt{8}}{\pi L}\left\vert
\frac{\mathfrak{P}\left(
\frac{\omega (A-a)}{L\sqrt{8}}\right) }{\mathfrak{P}^{\prime }\left( \frac{%
\omega (A-a)}{L\sqrt{8}}\right) }\right\vert ^{3}\Im m\frac{\left(
\mathfrak{P}^{\prime }\left( \frac{\omega (Z-a)}{L\sqrt{8}}\right) \mathfrak{%
P}^{\prime }\left( \frac{\omega (A-a)}{L\sqrt{8}}\right) \right) ^{2}}{%
\mathfrak{P}\left( \frac{\omega (Z-a)}{L\sqrt{8}}\right) \left( \mathfrak{P}%
^{\prime }\left( \frac{\omega (A-a)}{L\sqrt{8}}\right) \right) ^{2}-\left(
\mathfrak{P}^{\prime }\left( \frac{\omega (Z-a)}{L\sqrt{8}}\right) \right)
^{2}\mathfrak{P}\left( \frac{\omega (A-a)}{L\sqrt{8}}\right) }.
\eeqq
To finish, we shall provide the asymptotic for $d(Z;A)$ for large L.
Using the meromorphic property of $\mathfrak{P}$, we obtain the
following expansion of its derivative at the origin:
\beq
\mathfrak{P}'(\tilde Z)=-\frac{2}{\tilde Z^3}+O(1),
\eeq

\beq
\left\vert
\frac{\mathfrak{P}\left( \frac{\omega (A-a)}{L\sqrt{8}}\right) }{\mathfrak{P}^{\prime }\left( \frac{%
\omega (A-a)}{L\sqrt{8}}\right) }\right\vert^{3}\approx \left \vert\frac{\omega
A}{2L\sqrt{8}}\right\vert^3
\eeq
\beq
\left(
\mathfrak{P}^{\prime }\left( \frac{\omega (Z-a)}{L\sqrt{8}}\right) \mathfrak{%
P}^{\prime }\left( \frac{\omega (A-a)}{L\sqrt{8}}\right) \right)
^{2} \approx \frac{4(L\sqrt{8})^{12}}{(\omega^2 AZ)^6}
\eeq
\beq
\mathfrak{P}\left( \frac{\omega (Z-a)}{L\sqrt{8}}\right) \left( \mathfrak{P}%
^{\prime }\left( \frac{\omega (A-a)}{L\sqrt{8}}\right) \right)
^{2}\approx 2\left(\frac{L\sqrt{8}}{\omega Z}\right)^2
\left(\frac{L\sqrt{8}}{\omega A}\right)^6
\eeq
\beq
\left(
\mathfrak{P}^{\prime }\left( \frac{\omega (Z-a)}{L\sqrt{8}}\right) \right)
^{2}\mathfrak{P}\left( \frac{\omega (A-a)}{L\sqrt{8}}\right)\approx
2\left(\frac{L\sqrt{8}}{\omega A}\right)^2
\left(\frac{L\sqrt{8}}{\omega Z}\right)^6.
\eeq
Thus,
\beq
d(Z;A)&\approx& \frac{\omega \sqrt{8}}{\pi L}
\left \vert\frac{\omega
A}{2L\sqrt{8}}\right\vert^3 \frac{4(L\sqrt{8})^{12}}{(\omega^2)^6}
\frac{1}{2\left(\frac{L\sqrt{8}}{\omega}\right)^8}\Im m \left(
\frac{(AZ)^{-6}}{Z^{-2}A^{-6}-Z^{-6}A^{-2}}\right) \nonumber\\
       &\approx& \frac{16 \vert A \vert^3}{\pi} \Im m
       \left(\frac{(AZ)^{-6}}{Z^{-2}A^{-6}-Z^{-6}A^{-2}}\right) \nonumber \\
        &\approx& \frac{16 \vert A \vert^3}{\pi} \Im m
        \left(\frac{1}{Z^{4}-A^{4}}\right).
\eeq
\section*{The mean conditional time for a collision before exit}
We shall continue here with the expression for the mean conditional
time $\tau_{m}(\x)$ to meet before one of the particles escape. The
conditional time $\tau_{m}(\x)$ to hit the diagonal of the triangle
before the other sides is associated with the conditional process
$X^*$ solution of the stochastic differential equation
\cite{Karlin},
 \beqq dX^*(t)=2D\frac{\nabla p (X^*(t))}{ p
(X^*(t))}dt+\sqrt{2D}dW,
\eeqq
where $p$ is the probability (\ref{eqp}). $\tau_{m}(\x)$ satisfies
Dynkin's equation \cite{book}
\beq\label{eqfd1}
D p \Delta \tau_{m} +2D\nabla \tau_{m}\cdot \nabla p &=&-p \mbox{ in } T ,\nonumber\\
\tau_m &=&0 \mbox{ on } D,
\eeq
where D is the diagonal (there are no conditions on the other side).
Thus $w=p \tau_{m}$ satisfies:
\beq\label{eqfdd}
D  \Delta w &=&-p \mbox{ in } T ,\nonumber\\
w &=&0 \mbox{ on } \p T,
\eeq
We recall that the solution of the Dirichlet problem
\beq\label{eqfddd}
\Delta u &=&k \mbox{ in } H ,\nonumber\\
u &=&0 \mbox{ on } \p H,
\eeq
is
\beq\label{eqfdddd}
u(y) = \int_{H}G(y,z_1)k(z_1) dz_1
\eeq
where the Green function G is given by
\beq
G(y,z_1)= \frac{1}{2\pi} \ln\frac{|y-z_1|}{|y-\bar{z_1}|}.
\eeq
Using now that for any conformal transformation $\phi$,
\beq
\Delta (u o \phi) (z)=   |\phi'(z)|^2 \Delta u (\phi(z)) =-|\phi'(z)|^2
p(\phi(z)),
\eeq
we obtain
\beq
(u o \phi) (z) =-\frac{1}{D}\int_{H}|\phi'(z_1)|^2 p(\phi(z_1))
G(z-z_1) dz_1
\eeq
with $Z=\phi (z)=F(z)$,
\beq
u (Z) &=&-\frac{1}{D}\int_{H}|F'(z_1)|^2 p(F(z_1)) G(F^{-1}(Z),z_1) dz_1 \\
       &=&-\frac{1}{D}\int_{T}|F'|^2(F^{-1}(Z_1)) p(Z_1) G(F^{-1}(Z),F^{-1}(Z_1))
       \frac{dZ_1}{F'oF^{-1}(Z_1)}.
\eeq

\section*{Discussion}
The dynamics of double strand DNA (dsDNA) break is a fundamental
step of the repair process. There are no direct experimental
measurements yet of the dynamics of dsDNA ends (telomere) in the
confined nucleus environment, thus giving a fundamental role of the
theory in understanding the physics of motion, leaving aside the
molecular machinery involved in the repair process. Since the
general picture of telomere motions is not clear, we have presented
here a very coarse analysis based on Brownian motion, which can be
seen however as a drastic simplification of polymer motion in a
confined environment, restricted by the nuclear crowding, including
histones, the remaining DNA organization, nucleoli and many other
nuclear components. When the microdomain surrounding the dsDNA break
is sufficiently narrow (a long strip of length l), using the Rouse
model for the polymer, with a persistence length $l_0$, we can
distinguish two cases: $l<l_0$ or $l>l_0$. The polymer is modeled as
an ordered string of beads, each being connected to its next
neighbors by a spring of elasticity constant k. The mean length
between the beads is $l_0$. The motion of a string is governed by a
multi-dimensional Langevin equation, the potential of which is due
to the elastic forces. For a bead at position $x_i$, the motion is
described by the Smoluchowski limit of the Langevin equation
(i=1..N)
\beq
\dot{x}_i+\nabla U({x}_i,{x}_{i+1},{x}_{i-1}) =\sqrt{2D}\dot{w_i}
\eeq
where D is the diffusion constant, $w_i$ are $\delta-$correlated
Brownian motion of variance 1,  and the potential $U$ is
\beq
U({x}_{i-1},{x}_{i},{x}_{i+1})=
\left\{ \begin{array}{ccc}
U({x}_i,{x}_{i-1})+U({x}_i,{x}_{i+1}) &  \text{ if } $i=2..N-1$,\\
& &\\
 U({x}_j,{x}_{j+1}) &  \text{ if } j=1 \, or \, j=N-1
\end{array}\right.
\eeq
and
\beq
U({x}_i,{x}_{i+1})=k(\frac{1}{2}|x_{i+1}-x_{i}|^2-l_0
|x_{i+1}-x_{i}|)
\eeq
When $l<l_0$, the polymer cannot collapse and the DNA ends may be
approximated by the one dimensional motion of the polymer chain.
This approximation is so restrictive that the polymer relaxes to its
equilibrium and the two ends meet with probability one. When
$l>l_0$, there are two final possibilities: starting at an initial
position, either the two branches touch or curl up and then they
will not be able to be repaired in a reasonable time. We have
restricted our analysis to an one-dimensional Brownian motion. A
full analysis of this phenomenon is difficult and we shall discuss
now some ideas
 to address it. First our analysis is relevant for short DNA
fragments between two neighboring nucleosomes where we assimilate
the break location to the center of mass of the polymer, whose
motion is Brownian. However, the computation we presented here of
the probability to bind before escape cannot be generalized easily
to dimensions 2 or 3 because it depends heavily on conformal
mappings. To generalize our result, it is possible to use the Rouse
model for a polymer and estimate the probability that the two ends
of the dsDNA break meet for the first time before one of them
collapses. Similarly, estimating the probability that the two ends
meet before a given time would also be relevant. These questions are
much more difficult to address compared to our analysis. However, a
first step using simulations would be to estimate the mean first
passage time to one of the polymer end to reach a small hole. This
is already a nontrivial generalization of the small hole theory,
because the small hole is not small anymore. We hope that our
analysis will help to understand better the mechanisms of repair
processes occurring in the extremophilic bacterium Radiodurans,
where radiations are known to produce a nuclear phase transition,
leading to a restriction of the space and thus increasing the
probability of DNA repair \cite{NHEJ}.

{\noindent \bf Acknowledgements:} We thank N. Hoze for the Brownian
simulations and M. Yor for his interest and comments on this
manuscript. D.H. research's is supported by an ERC-starting grant.

\end{document}